\documentclass[sigconf]{acmart}

\usepackage{float}
\usepackage{stfloats}
\usepackage{booktabs}
\usepackage{enumitem}
\usepackage{amsmath}

\usepackage{amssymb}
\usepackage{multirow}

\AtBeginDocument{%
  }

\copyrightyear{2026}
\acmYear{2026}
\setcopyright{cc}
\setcctype{by}
\acmConference[SIGIR '26]{Proceedings of the 49th International ACM SIGIR Conference on Research and Development in Information Retrieval}{July 20--24, 2026}{Melbourne, VIC, Australia}
\acmBooktitle{Proceedings of the 49th International ACM SIGIR Conference on Research and Development in Information Retrieval (SIGIR '26), July 20--24, 2026, Melbourne, VIC, Australia}
\acmISBN{979-8-4007-2599-9/2026/07}
\acmDOI{10.1145/3805712.3808427}

\begin{document}

\title{Learning to Forget: Satiation-Aware Long-Sequence Transducers for Mitigating Post-Purchase Redundancy}

\author{Yipin Dai}
\affiliation{%
  \institution{Alibaba Group}
  \city{Hangzhou}
  \country{China}}
\email{220245072@seu.edu.cn}

\author{Ruocong Tang}
\authornote{Corresponding author.}
\affiliation{%
  \institution{Alibaba Group}
  \city{Hangzhou}
  \country{China}}
\email{tangruocong.trc@taobao.com}

\author{Xing Fang}
\affiliation{%
  \institution{Alibaba Group}
  \city{Hangzhou}
  \country{China}}
\email{fangxing.fx@taobao.com}

\author{Yang Huang}
\affiliation{%
  \institution{Alibaba Group}
  \city{Hangzhou}
  \country{China}}
\email{hy234680@taobao.com}

\author{Jing Wang}
\affiliation{%
  \institution{Alibaba Group}
  \city{Hangzhou}
  \country{China}}
\email{jing.wangj1@taobao.com}

\author{Zhentao Song}
\affiliation{%
  \institution{Alibaba Group}
  \city{Hangzhou}
  \country{China}}
\email{220245077@seu.edu.cn}

\author{He Guo}
\affiliation{%
  \institution{Alibaba Group}
  \city{Hangzhou}
  \country{China}}
\email{2401210268@stu.pku.edu.cn}

\renewcommand{\shortauthors}{Yipin Dai et al.}

\begin{abstract}
Sequential recommendation models predominantly interpret user interactions as positive signals for preference accumulation. However, in e-commerce scenarios, a purchase action often signifies the termination of a specific intent ("Interest Exit") rather than its continuation. Existing models overlook this distinction, suffering from Action-Intent Asymmetry, which leads to severe post-purchase redundancy. In this paper, we propose the Satiation-Aware Mechanism (SAM), an end-to-end framework designed to explicitly model the lifecycle of user interests. SAM incorporates three key components: (1) A Dual-path Cross-Attention architecture that retroactively suppresses historical clicks associated with a fulfilled intent while simultaneously retrieving personalized replenishment rhythms from long-term purchase history; (2) An Adaptive Satiation Gating Unit (ASGU) that generates a time-sensitive soft mask to inhibit satisfied interests immediately after purchase and gradually "re-awaken" them as the predicted repurchase cycle approaches; and (3) A self-supervised Time-to-Next-Purchase (TTNP) auxiliary task to learn latent product lifecycles without manual annotation. Extensive offline experiments on industrial datasets and online A/B testing demonstrate that SAM significantly reduces the Post-Purchase Repeat Rate (PPRR) by over 60\%. 
\end{abstract}



\maketitle

\section{Introduction}

Modern recommender systems and advertising platforms increasingly rely on long behavioral sequences to infer users' evolving intent. Transformer-style self-attention has become the dominant paradigm due to its capability to capture long-range dependencies and perform efficient parallel computation \cite{vaswani2017attention}. In sequential recommendation, models such as SASRec and BERT4Rec represent the state-of-the-art for next-item prediction \cite{kang2018sasrec, sun2019bert4rec}. Similarly, in industrial CTR/ranking scenarios, architectures like DIN and DIEN model candidate-aware activation and interest evolution over rich behavior histories, demonstrating production viability at scale \cite{zhou2018din, zhou2019dien}.

However, in e-commerce scenarios where purchase events are semantically critical, current systems overlook a fundamental phase in the user interest lifecycle: \textit{Interest Exit}. While standard attentive models excel at capturing interest evolution and accumulation, they fail to model the termination of an intent. This deficiency leads to what we term Action-Intent Asymmetry: a single purchase action often signifies the satisfaction and consequent closure of a broader underlying intent (manifested as a cluster of preceding clicks). Yet, standard self-attention mechanisms treat the purchase merely as another positive signal to be aggregated for future reinforcement \cite{li2020tisasrec}. This inability to recognize "Interest Exit" results in a persistent challenge known as \textit{post-purchase redundancy}, where fulfilled intents linger in the user representation, causing the system to recommend identical items immediately after conversion, thereby wasting premium recommendation slots and degrading user trust.

The difficulty in modeling Interest Exit lies in the inherent complexity of \textit{Satiation Dynamics}. The "exit" is rarely a binary event but a continuous process governed by product-specific lifecycles and user habits. First, the duration of satiation varies significantly: the purchase of a durable good (e.g., a refrigerator) typically signals a permanent or long-term exit of that intent, whereas a consumable (e.g., paper towels) implies a temporary exit with a latent, periodic resurrection need. Second, user consumption styles are diverse, ranging from "serial buying" to single-transaction satisfaction. Existing solutions often rely on post-hoc reranking rules (e.g., MMR \cite{carbonell1998mmr}, DPP \cite{kulesza2012dpp}) or heuristic probabilities between repeat and explore modes \cite{ren2019repeatnet}. While effective to some extent, these approaches operate outside the representation learning process and lack the granularity to model item-level satiation decay and the "lead-in" effects required for accurately predicting the re-emergence of an interest.

To tackle this issue, we propose the \textbf{Satiation-Aware Mechanism (SAM)}, an end-to-end framework designed to explicitly model the lifecycle of user interests. Unlike previous methods that treat interactions uniformly, SAM introduces a \textbf{Dual-path Cross-Attention} architecture to decouple intent localization from rhythm estimation. 
First, to address Action-Intent Asymmetry, we design a \textit{retroactive suppression} mechanism that uses the purchased item as a probe to identify and suppress not only the purchase event itself but also the relevant pre-purchase browsing history. 
Second, to capture personalized product lifecycles, we retrieve replenishment rhythms from a separate long-term purchase sequence ($S_{\text{buy}}$). 
These signals are fused into our proposed \textbf{Adaptive Satiation Gating Unit (ASGU)}. ASGU functions as a soft, time-sensitive mask within the attention layer: it suppresses attention to satisfied interests immediately after purchase and gradually "re-awakens" them as the elapsed time approaches the predicted cycle, controlled by a learnable lead-in offset.

A central challenge in SAM is learning accurate cycle embeddings without explicit manual labels (e.g., labeling items as "durable" or "consumable"). Drawing inspiration from continuous-time event modeling \cite{du2016rmtpp, mei2017hawkes}, we introduce a self-supervised auxiliary task: \textbf{Time-to-Next-Purchase (TTNP)}. By mining repurchase segments from raw logs and predicting the time gap to the next equivalent action, TTNP aligns the latent cycle embeddings with real-world repurchase rhythms, enabling ASGU to learn personalized "exit strength" across different users and categories.

In summary, our main contributions are:
\begin{itemize}
    \item \textbf{Satiation-Aware Paradigm:} We identify the Action-Intent Asymmetry problem and propose an end-to-end representation learning framework that treats purchase as a signal for intent termination rather than just accumulation.
    \item \textbf{Dual-path SAM Architecture:} We propose the Satiation-Aware Mechanism featuring a Dual-path structure. It combines retroactive suppression (to locate past exploratory clicks) with retrieval-based rhythm estimation (to model long-term habits).
    \item \textbf{Adaptive Gating \& TTNP Supervision:} We design the ASGU with a "lead-in" mechanism to balance redundancy suppression and repurchase recall, supervised by a Time-to-Next-Purchase auxiliary task that learns latent product lifecycles without manual annotation.
\end{itemize}

Empirically, SAM effectively mitigates post-purchase redundancy while improving long-term accuracy metrics on industrial-scale datasets, demonstrating that "knowing when to forget" is as crucial as "knowing what to remember." By implementing our framework, we enable models to autonomously determine when an interest thread should exit the active representation, effectively mitigating to the post-purchase redundancy problem.

\begin{figure*}[t]
  \centering
  \includegraphics[width=0.85\textwidth]{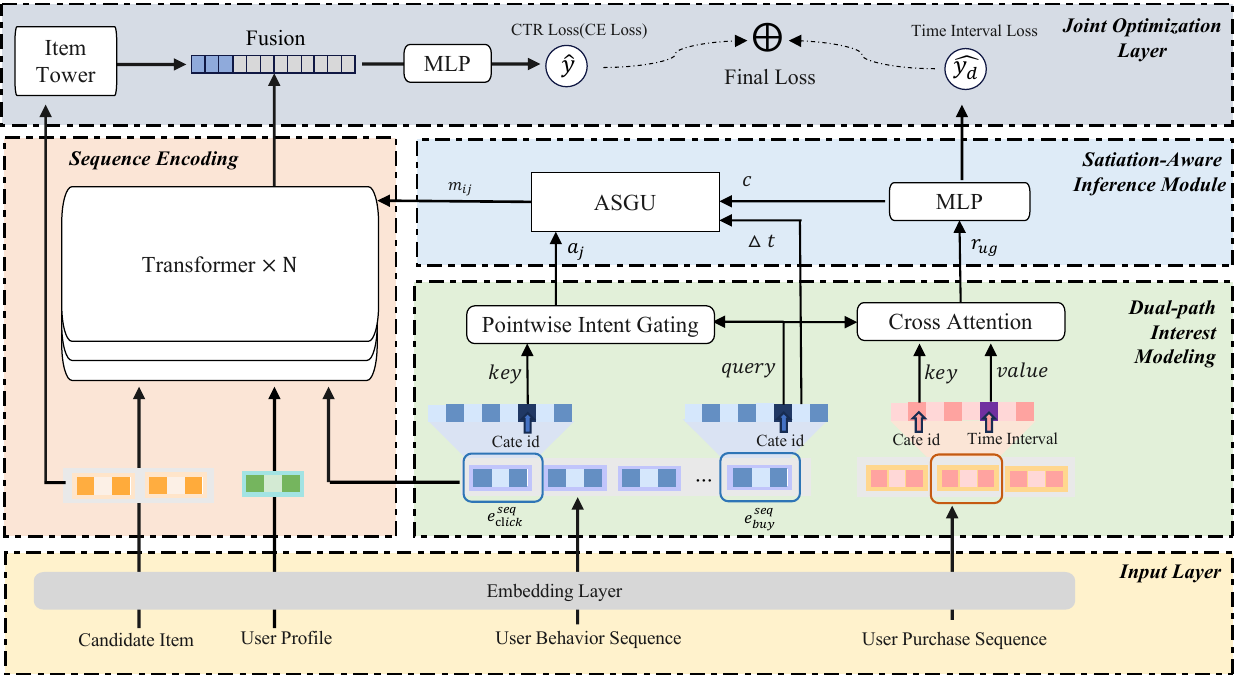}
  \caption{The overall architecture of the Satiation-Aware Mechanism (SAM).}
  \vspace{-4mm} 
  \label{fig:architecture}
\end{figure*}

\section{Methodology}
We propose the \textbf{Satiation-Aware Mechanism (SAM)} to explicitly model the \textit{Interest Exit} phenomenon in sequential recommendation. Unlike heuristic rule-based filtering, SAM learns personalized satiation cycles and retroactively suppresses historical intents in an end-to-end manner. The architecture consists of three core components: (1) a Dual-path Cross-Attention module for intent localization and rhythm retrieval, (2) an Adaptive Satiation Gating Unit (ASGU) for dynamic mask generation, and (3) a Satiation-aware Attention layer for logit-level intervention. As illustrated in Figure \ref{fig:architecture}, the SAM framework follows a multi-tower fusion architecture.

\subsection{Problem Formulation}
In the ranking stage of an industrial recommendation system, the objective is to estimate the Click-Through Rate (CTR) for a specific candidate item. Formally, given a user $u$, we define the inputs as follows: 1) \textbf{User Profile} $P_u$, which includes static features such as demographics and long-term preferences; 2) \textbf{User Behavior Sequence} $S = \{x_1, \dots, x_L\}$, a chronological sequence of clicks and purchases where each interaction $x_j$ contains item features and the category $g_j$; 3) \textbf{User Purchase Sequence} $S_{\text{buy}} = \{x^{(p)}_1, \dots, x^{(p)}_H\}$, a long-term history filtering only purchase events to capture consumption rhythms; and 4) \textbf{Candidate Item} $i$, the target item to be scored with category $g_i$.

The goal is to learn a scoring function $f(P_u, S, S_{\text{buy}}, i) \to \hat{y}$, where $\hat{y} \in [0, 1]$ represents the probability of interaction. Our specific focus is to correctly identify ``Interest Exit'' events within $S$ to suppress the score $\hat{y}$ when candidate $i$ belongs to a category the user has recently satisfied.

\subsection{Dual-path Interest Modeling}
To address the \textit{Action-Intent Asymmetry}, we employ a dual-path cross-attention mechanism that simultaneously identifies ``what to suppress'' and ``how long to suppress.'' To ensure semantic consistency and parameter efficiency, both paths share the same projection matrices ($W_Q, W_K$).

\subsubsection{Intent Localization (Retroactive Suppression)}
A purchase signals the fulfillment of an entire intent cluster. To avoid the weight dilution inherent in Softmax for long sequences (e.g., $L=500$), we define the intent attribution weight $a_j$ as a pointwise similarity gate. We use the category embedding $\mathbf{e}_{g_{\text{buy}}}$ of the purchase as a probe query against the category embeddings $\mathbf{e}_{g_j}$ of historical nodes:
\begin{equation}
a_j = \sigma \left( \frac{(W_Q \mathbf{e}_{g_{\text{buy}}})^\top (W_K \mathbf{e}_{g_j})}{\sqrt{d}} \right) \tag{1}
\end{equation}
where $\sigma$ is the Sigmoid function. Unlike Softmax, this pointwise formulation allows $a_j$ to maintain a high suppression strength (near 1.0) for all nodes within the same intent cluster, regardless of sequence length.

\subsubsection{Personalized Rhythm Retrieval}
Simultaneously, we retrieve the user's consumption rhythm from the long-term purchase history $S_{\text{buy}} = \{ (\mathbf{e}_{g_k}, \mathbf{e}_{\Delta T_k}) \}_{k=1}^H$. We employ target-aware cross-attention to compute a rhythm representation $\mathbf{r}_{u,g}$:
\begin{equation}
\mathbf{r}_{u,g} = \sum_{k=1}^H \beta_{k} \cdot (W_V \mathbf{e}_{\Delta T_k}), \quad \beta_k = \mathrm{Softmax}_k \left( \frac{(W_Q \mathbf{e}_{g_{\text{buy}}})^\top (W_K \mathbf{e}_{g_k})}{\sqrt{d}} \right) \tag{2}
\end{equation}

where $\mathbf{e}_{\Delta T_k}$ is the embedding of historical time intervals. The resulting $\mathbf{r}_{u,g}$ is a weighted sum of historical paces, which is then decoded by an MLP into a positive scalar $c = \mathrm{Softplus}(\mathrm{MLP}(\mathbf{r}_{u,g}))$, representing the predicted replenishment cycle in days. This cycle $c$ is explicitly supervised by the TTNP auxiliary task (detailed in Section 2.5). The shared $W_Q, W_K$ ensure that the rhythm retrieval and intent localization paths operate within a unified semantic space.

\begin{table*}[h]
\centering
\caption{Offline and Online comparison results. (Mean $\pm$ Std.) * indicates $p < 0.05$. In the Online A/B section, all metrics are reported as relative percentage improvements/reductions over the deployed baseline (DIN).}
\label{tab:main_results}
\begin{tabular}{l|ccc|ccc}
\toprule
\multirow{2}{*}{\textbf{Model}} & \multicolumn{3}{c|}{\textbf{Offline Metrics}} & \multicolumn{3}{c}{\textbf{Online A/B Testing}} \\
 & \textbf{AUC ($\uparrow$)} & \textbf{GAUC ($\uparrow$)} & \textbf{PPRR ($\downarrow$)} & \textbf{$\Delta$ CTR ($\uparrow$)} & \textbf{$\Delta$ GMV ($\uparrow$)} & \textbf{$\Delta$ Bad-Case ($\downarrow$)} \\ \midrule
\textit{Classic Models} & & & & & & \\
DNN & 0.741 $\pm$ 0.001 & 0.665 $\pm$ 0.002 & 10.5\% & - & - & - \\
DCN-v2 \cite{wang2021dcnv2} & 0.745 $\pm$ 0.001 & 0.669 $\pm$ 0.001 & 11.2\% & - & - & - \\
DIN \cite{zhou2018din} (Baseline) & 0.749 $\pm$ 0.001 & 0.674 $\pm$ 0.001 & 15.5\% & 0.00\% & 0.00\% & 0.00\% \\ 
DIN + Fixed-Window Filter & 0.745 $\pm$ 0.001 & 0.670 $\pm$ 0.002 & 7.7\% & -0.2\% & -0.3\% & -31.9\% \\ \midrule
\textit{Recent SOTA} & & & & & & \\
HiFormer \cite{gui2023hiformer} & 0.753 $\pm$ 0.002 & 0.679 $\pm$ 0.002 & 14.5\% & - & - & - \\
RankMixer \cite{zhu2025rankmixer} & 0.754 $\pm$ 0.001 & 0.680 $\pm$ 0.001 & 13.8\% & - & - & - \\
HSTU \cite{zhai2024actions} & 0.758 $\pm$ 0.001 & 0.684 $\pm$ 0.001 & 15.2\% & +0.8\% & +0.4\% & +11.0\% \\
HHFT \cite{zhang2025hhft} & 0.759 $\pm$ 0.002 & 0.685 $\pm$ 0.002 & 14.9\% & - & - & - \\
MTGR \cite{han2025mtgr} & 0.761 $\pm$ 0.001 & 0.687 $\pm$ 0.001 & 16.1\% & - & - & - \\
OneTrans \cite{zhang2026onetrans} & \underline{0.762 $\pm$ 0.001} & \underline{0.689 $\pm$ 0.001} & 15.8\% & +1.1\% & +0.6\% & +10.2\% \\ \midrule
\textbf{SAM (Ours)} & \textbf{0.763$^*$ $\pm$ 0.001} & \textbf{0.691$^*$ $\pm$ 0.001} & \textbf{4.1\%} & \textbf{+1.1\%} & \textbf{+0.9\%} & \textbf{-74.5\%} \\ \bottomrule
\end{tabular}
\end{table*}

\subsection{Adaptive Satiation Gating Unit (ASGU)}
The ASGU generates a satiation mask $m_{ij}$ to modulate historical nodes. To balance immediate redundancy suppression with future demand forecasting, we design the mask to be deepest right after purchase and gradually release as it approaches the predicted cycle $c$:
\begin{equation}
m_{ij} = 1 - a_j \cdot \underbrace{\sigma \left( k \cdot \left( \alpha - \frac{\Delta t_{ij}}{c} \right) \right)}_{\text{Satiety Function } \phi} \tag{3}
\end{equation}
where $\Delta t_{ij}$ is the elapsed time since purchase and $k$ is a learnable steepness factor. 

Specifically, in the \textbf{Satiety Phase}, when $\Delta t_{ij} \to 0$, the term $(\alpha - \frac{\Delta t}{c}) \approx \alpha$. Given $\alpha \in (0,1]$ is a Lead-in Offset (e.g., 0.8), $\phi$ remains high, leading to $m_{ij} \approx 1 - a_j$. For identical categories ($a_j \approx 1$), the node is effectively masked ($m_{ij} \to 0$). During the \textbf{Recovery Phase}, as $\Delta t_{ij}$ approaches the threshold $\alpha c$, the term inside $\sigma$ nears 0, causing $\phi \to 0.5$ and the mask to fade. Finally, for \textbf{Pre-purchase Recall}, once $\Delta t_{ij} > \alpha c$, $\phi$ drops towards 0, and $m_{ij} \to 1$. This ``lead-in'' mechanism ensures that the model can re-activate historical interests slightly before the predicted cycle $c$ ends to capture early browsing for replenishment.

\subsection{Satiation-aware Attention}
To maintain numerical stability, we inject the gating signal as a Logit-bias Mask. For any query $i$ and key $j$, the modified attention logit $\ell_{ij}$ is formulated as:
\begin{equation}
\ell_{ij} = \frac{\mathbf{q}_i^\top \mathbf{k}_j}{\sqrt{d}} + \log( \mathrm{clip}(m_{ij}, \epsilon, 1) ) \tag{4}
\end{equation}
This ensure that when $m_{ij} \to 0$, the corresponding attention weight $\alpha_{ij} \to 0$.

\subsection{TTNP Auxiliary Task and Training}
To ground the latent cycle $c$ in physical time, we introduce the Time-to-Next-Purchase (TTNP) auxiliary task. For a sequence of purchases of the same category $g$ at times $t_k$ and $t_{k+1}$, the model minimizes a censored regression loss:
\begin{equation}
\mathcal{L}_{\text{aux}} = \sum \left| \log(c + 1) - \log(\Delta t + 1) \right|^2 \tag{5}
\end{equation}
where $\Delta t = t_{k+1} - t_k$ is the target interval to the next purchase. Crucially, to handle right-censored cases where a user only has a single purchase of category $g$ without a subsequent purchase in our observation window, we impute the missing $\Delta t$ using the platform's global average repurchase cycle $\bar{c}_g$ for that category. Specifically, we set $\Delta t = \max(T_{\text{end}} - t_k, \bar{c}_g)$, where $T_{\text{end}}$ is the end of the data window. This ensures the model learns a robust lower bound even for sparse, long-cycle durables.

The total loss is a joint optimization: $\mathcal{L} = \mathcal{L}_{\text{CTR}} + \lambda \mathcal{L}_{\text{aux}}$.

\section{Experiments}

\subsection{Experimental Setup}

\subsubsection{Datasets}
We collected and sampled the online service logs from the online Tmall e-commerce platform between 2025/06/07 to 2025/11/07 to construct the experimental dataset $\mathcal{D}$. As shown in Table \ref{tab:dataset_stats}, the entire dataset is split chronologically into a training set $\mathcal{D}_{\text{train}}$ (first 5 months) and a testing set $\mathcal{D}_{\text{test}}$ (last 1 month) to prevent data leakage. We filter for active users with at least 20 interactions and truncate sequences to length $L=500$.

\begin{table}[H]
\centering
\caption{Statistics of the established Industrial Logs.}
\label{tab:dataset_stats}
\begin{tabular}{lccccc}
\toprule
\textbf{Dataset} & \textbf{\#Users} & \textbf{\#Items} & \textbf{\#Actions} & \textbf{\#Purchases} \\ \midrule
$\mathcal{D}_{\text{train}}$ & 11.52M & 7.82M & 190.33M & 9.06M  \\
$\mathcal{D}_{\text{test}}$ & 6.13M & 4.65M & 52.74M & 3.22M  \\ \bottomrule
\end{tabular}
\end{table}

\subsubsection{Evaluation Metrics}
Since we focus on the fine-ranking stage (CTR prediction), we adopt a comprehensive set of metrics to evaluate both accuracy and diversity. For \textbf{Ranking Accuracy}, we utilize \textbf{AUC} (Area Under ROC Curve) to measure global ranking capability and \textbf{GAUC} (Group AUC) to assess quality within each user's personalized list, the latter being more indicative of online performance in industrial systems \cite{zhou2018din}. Beyond accuracy, we introduce \textbf{Redundancy Control} metrics to quantify the "Interest Exit" capability: \textbf{PPRR} (Post-Purchase Repeat Rate) tracks the global percentage of impressions where the recommended item shares the exact same leaf-category as the user's most recent purchase, computed within a category-specific time window drawn from a pre-existing repurchase cycle lookup table \cite{carbonell1998mmr, kulesza2012dpp}. Meanwhile, the \textbf{Bad-Case Ratio} is a direct online business metric: it measures the proportion of "already purchased" reasons among all explicit negative feedback events (e.g., clicking "dislike") submitted by users. For online A/B testing, we monitor \textbf{CTR} and \textbf{GMV} on complementary items to ensure balanced business growth.

\subsubsection{Competitors}
We compare SAM against a comprehensive set of baselines categorized into three groups. The first group consists of \textbf{Classic \& Interaction Models}, including standard \textbf{DNN} \cite{cheng2016wide, covington2016deep}, \textbf{DCN-v2} \cite{wang2021dcnv2}, and \textbf{DIN} \cite{zhou2018din}, which focus on feature interactions and candidate-relevant user interests. The second group covers \textbf{Efficient Sequence Models} such as \textbf{HiFormer} \cite{gui2023hiformer} and \textbf{RankMixer} \cite{zhu2025rankmixer}, utilizing hierarchical mechanisms for efficiency. The third group represents \textbf{Generative/Large-Scale Models}, including \textbf{HSTU} \cite{zhai2024actions}, \textbf{MTGR} \cite{han2025mtgr}, \textbf{OneTrans} \cite{zhang2026onetrans}, and \textbf{HHFT} \cite{zhang2025hhft}, which are state-of-the-art architectures designed for extremely long-term history modeling.

\subsection{Experimental Results}

\subsubsection{SOTA Comparison}
For offline evaluation, all experiments were repeated 5 times on $\mathcal{D}_{\text{test}}$. For online A/B testing (following standard protocols \cite{kohavi2009controlled}), models are deployed to serve 1\% of live traffic. The comparative results are presented in Table \ref{tab:main_results}.

Analyzing the results reveals a clear evolution in model performance. Traditional models like DNN and DCN-v2 show limited GAUC as they fail to capture sequential dependencies. While DIN significantly improves accuracy by introducing attention mechanisms, it inadvertently exacerbates redundancy because it explicitly attends to the purchased item if the candidate is similar. Modern SOTA sequence models (e.g., MTGR, OneTrans) achieve the highest GAUC scores, proving their ability to capture long-term interests; however, their high PPRR indicates a critical flaw—they treat post-purchase signals merely as ``strong positive feedback,'' leading to severe post-purchase redundancy. 

In contrast, our \textbf{SAM} framework achieves a significant improvements across all metrics. It statistically ties with or surpasses OneTrans in AUC and GAUC, while crucially reducing the \textbf{PPRR} by over \textbf{60\%} compared to the strongest baseline. This confirms that SAM correctly "exits" satisfied interests without compromising the prediction accuracy for other relevant items. Online A/B testing further validates this impact, showing a \textbf{+1.1\%} lift in CTR and a \textbf{+0.9\%} lift in GMV, driven by the effective reduction of redundant exposures which allows users to explore new categories. Furthermore, a quantitative breakdown confirms that SAM successfully captures distinct product lifecycles: it predicts extended satiation cycles averaging over 300 days for typical durables (e.g., refrigerators), while correctly inferring much shorter rhythms ($\sim$15--30 days) for high-frequency consumables (e.g., tissues). By learning these personalized cycles rather than relying on a rigid heuristic baseline (e.g., \textit{DIN + Fixed-Window Filter}, which severely degrades GAUC to 0.670 and drops online GMV by 0.3\% due to indiscriminate over-suppression), SAM dynamically balances redundancy mitigation with legitimate repurchase demands.

\subsubsection{Ablation Study}
To validate the contribution of each component, we conduct an ablation study shown in Table \ref{tab:ablation}. The observed order of contribution to GAUC and Redundancy reduction is \textbf{Dual-path Retrieval > TTNP > ASGU Gating}. Specifically, removing the rhythm retrieval module causes a slight drop in GAUC but a sharp increase in PPRR, confirming that historical purchase rhythm is the primary signal for determining "when to stop." Furthermore, without the auxiliary TTNP loss, the model fails to learn the diverse satiation cycles of different goods (e.g., distinguishing refrigerators from tissues), leading to sub-optimal GAUC. Finally, replacing the adaptive ASGU with a hard mask maintains low redundancy but hurts GAUC, as it mistakenly suppresses "lead-in" interests such as pre-repurchase browsing.

\begin{table}[H]
\centering
\caption{Ablation outcomes on SAM variants. GAUC is the primary accuracy metric.}
\label{tab:ablation}
\begin{tabular}{lcc}
\toprule
\textbf{Variant} & \textbf{GAUC ($\uparrow$)} & \textbf{PPRR ($\downarrow$)} \\ \midrule
\textbf{Full SAM} & \textbf{0.691} & \textbf{4.1\%} \\
w/o TTNP Task & 0.688 ($\downarrow$) & 8.4\% ($\uparrow$) \\
w/o Dual-path Retrieval & 0.682 ($\downarrow$) & 12.9\% ($\uparrow\uparrow$) \\
w/o ASGU (Hard Mask) & 0.685 ($\downarrow$) & 4.0\% ($\approx$) \\ \bottomrule
\end{tabular}
\end{table}

\section{Conclusion}
In this paper, we propose the \textbf{Satiation-Aware Mechanism (SAM)}, a novel framework to address the post-purchase redundancy problem in industrial fine-ranking systems. By introducing a Dual-path Context Retrieval module and an adaptive gating unit supervised by a Time-to-Next-Purchase (TTNP) task, SAM effectively models the ``Interest Exit'' phenomenon.
Extensive experiments on industrial datasets demonstrate that SAM significantly reduces redundancy metrics (PPRR) while maintaining state-of-the-art ranking accuracy (GAUC). Online A/B testing further validates its business value, showing substantial gains in CTR and GMV. These findings highlight the importance of explicitly modeling intent termination in long-sequence user behavior learning.

\section{Acknowledgments}
We acknowledge the Alibaba PAI Team for releasing TorchEasyRec\cite{torcheasyrec2024}, which supported the engineering implementation and large-scale training of our experiments.

\bibliographystyle{ACM-Reference-Format}
\balance
\bibliography{sample-base}


\end{document}